\documentclass[12pt]{iopart}

\usepackage{graphics}

\begin{document}
\title{Coriolis mixing of the K=1 and K=0 mixed symmetry states in the well deformed even-even nuclei.}

\author{N. Yu. Shirikova,  A. V. Sushkov},
\address{Joint Institute for Nuclear Research, 141980 Dubna, Moscow region, Russia}
\author{ R. V. Jolos} 
\address{Joint Institute for Nuclear Research and Dubna State University, 141980 Dubna, Moscow region, Russia}
\ead{jolos@theor.jinr.ru}
\vspace{10pt}
\begin{indented}
\item[]March 2022
\end{indented}

\begin{abstract}
The Coriolis matrix elements responsible for mixing of the $1^+ K=1$ and $1^+ K=0$ states are calculated in the framework of the Quasiparticle Phonon Model for several Gd and Dy isotopes. In many considered cases these matrix elements are equal to several tens of keV and are comparable with energy distances between the mixed levels. The results obtained indicates that Gd isotopes could be more suitable for finding deviations from Alaga rules in M1 transitions from $1^+$ state to the states of the ground band.
\end{abstract}

%
%
%
%

\section{Introduction}
\label{intro}
The lowest lying states of the axially symmetric deformed nuclei are related to rotational degrees of freedom.
The symmetries of the deformed mean field of atomic nuclei determine the possible sets of quantum numbers
characterizing ground and excited states. Example of such symmetry is the invariance with respect to a rotation of 180$^{\circ}$ around the arbitrary chosen axis perpendicular to the symmetry axis (${\cal R}$ symmetry) \cite{BM2}. As a consequence of this symmetry, the rotational bands with
$K^{\pi}=0^+$ comprise only states with even ($r$=+1) or odd ($r$=--1) angular momenta, where $r$ is ${\cal R}$ symmetry quantum number. The sequence of the angular momenta for $r$=+1 corresponds to the well known ground state rotational bands of axially symmetric deformed nuclei: $0^+, 2^+, 4^+$, ... The sequence of the angular momenta for $r$=--1 should start with the $1^+$ state. However, information on the states with
$I^{\pi}K=1^+ K=0$ is steel practically absent. The exception is the result obtained in~\cite{Beck1}, 
where $K$ - mixing of the close-lying $K=1$ and $K=0$ states was observed. The $I^{\pi}K=1^+ K=1$ states,
which are characterized by the strong M1 transition probabilities to the ground states are known under the name "`mixed symmetry"' states. They are well investigated and it is known that they are quite fragmented~\cite{Bohle1}.


In the Q-phonon scheme of IBM-2~\cite{Siems,Otsuka1,Pietralla2} for the mixed symmetry states the wave function of the $1^+$ state can be written in a compact notation as
\begin{eqnarray}
\label{Eq1}
|1^+_{ms}\rangle=\left(Q_sQ_m\right)_1|0^+_1\rangle.
\end{eqnarray}
Here $Q_s=Q_{\pi}+Q_{\nu}$ denotes $F$-scalar and $Q_m=Q_{\pi}/N_{\pi}-Q_{\nu}/N_{\nu}$ 
denotes mixed-symmetric quadrupole operators. The $Q$-phonon expression (\ref{Eq1}) represents a good approximation
over large range of values of the parameters of the IBM-2 Hamiltonian \cite{Kim1}. In the limit of the well
deformed nuclei
\begin{eqnarray}
\label{Eq2}
|1^+_{ms}\rangle=\sum_{K,k,k'}D^1_{MK}C^{1K}_{2k 2k'}Q_{s,k}Q_{m,k'},
\end{eqnarray}
where $D^1_{MK}$ is the Wigner function. For the state with $K$=0 we obtain that
\begin{eqnarray}
\label{Eq3}
|1^+_{ms}\rangle=D^1_{M0}\left(C^{10}_{20  20}Q_{s,0}Q_{m,0}-C^{10}_{22  2-2}Q_{s,2}(Q_{m,2}-Q_{m,-2})\right).
\end{eqnarray}
We see from (\ref{Eq3}) that since $C^{10}_{20  20}=0$ the state
 $1^+$ with $K=0$  can be only two-phonon (i.e. four-quasiparticle) state in the collective model approach
with mixed-symmetry phonon coupled to the symmetric $\gamma$-phonon. The two-phonon structure of the
$1^+ K=0$ state leads to a weak M1 transition from the ground state to this $1^+$ mixed symmetry state.
This fact explains why it is so difficult to detect these states. At the same time, the $1^+ K=1$ state has a one-phonon component as the main one. We mention, that in contrast to the case of $1^+ K=0$ state, the collective
$1^- K=0$ state can be realized as a one-phonon state.

In the dynamical symmetry scheme of IBM-2 several $I^{\pi}=1^+$ states are existed in (N-1,1) sector
of the mixed-symmetry states. However, they are characterized by the different total numbers of quadrupole bosons.
Only one of these states is characterized by the strong B(M1) value for transition to the ground state.

The mixing of the lowest IBM-2  $I^{\pi}=1^+$ mixed-symmetry states with the other $I^{\pi}=1^+$
states has been considered in \cite{Scholten} where the numerical value for an average mixing matrix element of $\approx 100$ keV was obtained.  It is also pointed out in \cite{Scholten} that for the wave functions corresponding to the SU(3) limit of IBM-2 the mixing matrix element can become larger than in the vibrational limit.

Let us consider a structure of the $1^+$ mixed-symmetry states from the microscopic model point of view.
The single particle states $\varphi_K(q)$ of the axially deformed nuclei are presented by the following expansion in the spherical shell model basis
\begin{eqnarray}
\label{Eq4}
\varphi_K(q)=\sum_{l,j}d^q_{lj}|NljK\rangle\equiv |q\sigma\rangle,
\end{eqnarray}
where $\sigma=\pm 1$ marked states conjugate with respect to the time reflection operation
\begin{eqnarray}
\label{Eq5}
{\cal T}|q+\rangle =|q-\rangle =(-1)^{K+1/2}\varphi_{-K}(q),
\end {eqnarray}
where
\begin{eqnarray}
\label{Eq6}
\varphi_{-K}(q)=\sum_{l,j}d^q_{lj}(-1)^{l+j-1/2}|Nlj-K\rangle.
\end {eqnarray}
The two-quasiparticle operators used to construct excited state vectors in even-even deformed nuclei
are divided \cite{Soloviev1} into electric type
\begin{eqnarray}
\label{Eq7}
A^+_{qq'}=\frac{1}{\sqrt{2}}\sum_{\sigma}\sigma\alpha^+_{q,-\sigma}\alpha^+_{q'\sigma},\nonumber\\
{\bar A}^+_{qq'}=\frac{1}{\sqrt{2}}\sum_{\sigma}\alpha^+_{q',\sigma}\alpha^+_{q,\sigma},
\end{eqnarray}
and magnetic type
\begin{eqnarray}
\label{Eq8}
{\cal U}^+_{qq'}=\frac{1}{\sqrt{2}}\sum_{\sigma}\alpha^+_{q,-\sigma}\alpha^+_{q'\sigma},\nonumber\\
{\bar {\cal U}}^+_{qq'}=\frac{1}{\sqrt{2}}\sum_{\sigma}\sigma\alpha^+_{q',\sigma}\alpha^+_{q,\sigma},
\end{eqnarray}
operators, which are used for construction of the electric transition and magnetic transition operators, correspondingly. The so called mixed symmetry states which are characterized by the strong M1
transitions from the ground state and weak E2 transitions to the symmetric states are constructed using the
magnetic type two-quasiparticle creation operators ${\cal U}^+_{qq'}$ and ${\bar{\cal U}}^+_{qq'}$. We mention,
that in the case of absence of the residual interaction two-quasiparticle states $A^+_{q_1q_2}|0^+_1\rangle$
and ${\cal U}^+_{q_1q_2}|0^+_1\rangle$ with the same $q_1, q_2$ have the same excitation energies.
However, they are characterized by different electromagnetic transition properties. Of course, residual forces  create a difference in their excitation energies.

We set that in the microscopic approach, when not only collective but many other two-quasiparticle states
are considered, many $1^+$ states with both $K$=1 and $K$=0 can be constructed. For instance,
two-quasiparticle state of the magnetic type
\begin{eqnarray}
\label{Eq10}
{\cal U}^+_{qq}=\frac{1}{\sqrt{2}}\sum_{\sigma}\alpha^+_{q,-\sigma}\alpha^+_{q\sigma}|0^+_1\rangle
\end{eqnarray}
has $K$=0.

The experimental information on the M1 excitations with $K$=0 projection was missing so far
for axially deformed even-even nuclei. Only recently such information on the M1 excitation strength
and decay characteristics of the $I^{\pi}K=1^+K=0$ state with excitation energy around 3 MeV
was obtained for $^{164}$Dy \cite{Beck1}. It was found for the two observed 1$^+$ states that the branching ratios deviate significantly from Alaga rule. This deviation can only be explained by mixing of the underlying basis
states with projection quantum numbers $K=0$ and $K=1$.

It is the aim of this paper to calculate the value of the Coriolis mixing matrix element for the
$1^+$ states with $K$=0 and $K$=1  for the well deformed axially symmetric rare earth nuclei. In the case when these mixing matrix elements are not small compare to the energy distances between the mixed levels this might indicate the  existance of the $1^+$ states with a decay behavior deviating from the Alaga predictions.
The calculations are performed below basing on the Quasiparticle-Phonon Model (QPNM) \cite{Soloviev1}.

\section{Brief description of the Quasiparticle-Phonon Model (QPNM)}

The intrinsic Hamiltonian of the QPNM has the following structure
\begin{eqnarray}
\label{Eq11}
H=H_{sp}+H_{pair}+H_{ph}+H_{pp}.
\end{eqnarray}
The first term in (\ref{Eq11}) is the proton and neutron single particle Hamiltonian. The single particle
energies are obtained from a deformed axially symmetric Woods-Saxon potential
\begin{eqnarray}
\label{Eq12}
V_{sp}=V(r)+V_{ls}(r),
\end{eqnarray}
\begin{eqnarray}
\label{Eq13}
V(r)=\frac{-V_0^{\tau}}{1+\exp\left(\alpha(r-R(\theta,\varphi))\right)},\\
V_{ls}=-\kappa{\vec p}\otimes{\vec \sigma}\cdot\nabla V(r),
\end{eqnarray}
where ($\tau$=p,n)
and
\begin{eqnarray}
\label{Eq15}
R(\theta,\varphi)=R_0\left(1+\beta_2Y_{20}(\theta,\varphi)+\beta_4Y_{40}(\theta,\varphi)\right).
\end{eqnarray}
Here $R_0=r_0A^{1/3}$ is the radius of the spherical nucleus having the same size; $\beta_2$
and $\beta_4$ are the quadrupole and hexadecapole axial deformation parameters.

The second term in (\ref{Eq11}) is a proton and neutron monopole pairing interaction. The terms
$H_{ph}$ and $H_{pp}$ are separable interactions acting, respectively, in the particle-hole (ph) and
particle-particle (pp) channels. The ph term is composed of spin-independent and spin-dependent parts
\begin{eqnarray}
\label{Eq16}
H_{ph}=H_M+H_S,
\end{eqnarray}
with
\begin{eqnarray}
\label{Eq17}
H_M=-\frac{1}{2}\sum_{\tau=p,n}\sum_{\rho=\pm 1}\sum_{\lambda\mu\sigma}\left(\kappa_0^{\lambda\mu}+\rho\kappa_1^{\lambda\mu}\right)M^+_{\lambda\mu\sigma}(\tau)M_{\lambda\mu\sigma}(\rho\tau),\\
H_S=-\frac{1}{2}\sum_{\tau=p,n}\sum_{\rho=\pm 1}\sum_{l\lambda\mu\sigma}\left(\kappa_0^{l\lambda\mu}+\rho\kappa_1^{l\lambda\mu}\right)S^+_{l\lambda\mu\sigma}(\tau)S_{l\lambda\mu\sigma}(\rho\tau).
\end{eqnarray}
Above, if $\tau =p(n)$  then --$\tau =n(p)$.
Here
\begin{eqnarray}
\label{Eq19}
M^+_{\lambda\mu\sigma}=\sum_{q_1\sigma_1 q_2\sigma_2}\langle q_1\sigma_1|R_{\lambda}(r)Y_{\lambda\sigma\mu}(\vartheta\varphi)|q_2\sigma_2\rangle a^+_{q_1\sigma_1}a_{q_2\sigma_2},\\
S^+_{l\lambda\mu\sigma}\sum_{q_1\sigma_1 q_2\sigma_2}\langle q_1\sigma_1|R_{\lambda}(r)[{\vec \sigma}\otimes Y_l]_{\lambda\sigma\mu}|q_2\sigma_2\rangle a^+_{q_1\sigma_1}a_{q_2\sigma_2}.
\end{eqnarray}
The symbols $\kappa_0^{\lambda\mu}$ and $\kappa_1^{\lambda\mu}$ denote, respectively, the isoscalar
and isovector interaction constants of the spin-independent ($\lambda, \mu$) terms. The spin-dependent
analogues are  $\kappa_0^{l\lambda\mu}$ and $\kappa_1^{l\lambda\mu}$.

The particle-particle term is spin-independent and has the form
\begin{eqnarray}
\label{Eq21}
H_{pp}=-\frac{1}{2}\sum_{\tau=p,n}\sum_{\lambda\mu\sigma}G_{\tau}^{\lambda\mu}P^+_{\lambda\mu\sigma}(\tau)P_{\lambda\mu\sigma}(\tau),
\end{eqnarray}
where
\begin{eqnarray}
\label{Eq22}
P^+_{\lambda\mu\sigma}=\sum_{q_1\sigma_1 q_2\sigma_2}\langle q_1\sigma_1|R_{\lambda}(r)Y_{\lambda\sigma\mu}(\vartheta\varphi)|q_2\sigma_2\rangle \sigma_2 a^+_{q_1\sigma_1}a^+_{q_2\sigma_2}.
\end{eqnarray}
The quantities $G_{\tau}^{\lambda\mu}$ are the pairing strength constants. In all ph and pp terms the
radial factor is $R_{\lambda}(r)=dV(r)/dr$, where $V(r)$ is the central part of the Woods-Saxon potential
(\ref{Eq13}).

The next step is to express the Hamiltonian in terms of the quasiparticle operators $\alpha^+_{q\sigma}$,
$\alpha_{q\sigma}$ and then to express the RPA phonon operators in terms of the two-quasiparticle creation and annihilation operators (\ref{Eq7}-\ref{Eq8}). Details of definition of the Hamiltonian parameters are given in \cite{Soloviev2}. The single particle spectrum used in the calculations is taken from the bottom of the potential and up to +5 MeV.  The two-quasiparticle configurations with the energies  up to 30 MeV
are included into calculations. Because of the separable form of the interaction not all the terms of the Hamiltonian contribute to the structure of the given RPA phonon operator. The quadrupole-quadrupole interaction is the most important ingredient of the interaction part of the Hamiltonian. The spin-spin part has the effect of pushing the spin excitations upward in energy thereby enforcing the orbital character of the low-lying states. Spin-multipole and tensor interaction have negligible effects.

The matrix elements of the Coriolis interaction
\begin{eqnarray}
\label{Eq24}
H_{Coriolis}=-\frac{\hbar^2}{2\Im}\left(I_+j_- + I_-j_+\right),
\end{eqnarray}
where $I_{\pm}$ are components of the total angular momentum operator, acting on the Wigner functions 
$D^I_{M K}$, and $j_{\pm}$ are components of the intrinsic angular momentum operator,
are calculated using the following expressions for the wave vectors of the $I^{\pi}K=1^+1$ and $1^+0$ states:
\begin{eqnarray}
\label{Eq25}
|1^+MK=1\rangle =\sqrt{\frac{3}{16\pi^2}}\left(D^1_{M 1}Q^+(K=1,\sigma=+1)\right.\nonumber\\
\left. + D^1_{M -1}Q^+(K=1,\sigma=-1)\right)|0\rangle ,
\end{eqnarray}
where the phonon creation operators $Q^+(K,\sigma)$ are
\begin{eqnarray}
\label{Eq26}
Q^+(K=1,\sigma)=\frac{1}{2\sqrt{2}}\sum_{q_1 q_2}\psi^{1^+,K=1}_{q_1 q_2}\left({\cal U}^+_{q_1 q_2}\cdot\sigma(k_{q_1}-k_{q_2})-A^+_{q_1 q_2}\right),
\end{eqnarray}
and
\begin{eqnarray}
\label{Eq27}
|1^+MK=0\rangle =\sqrt{\frac{3}{8\pi^2}}D^1_{M 0}\sum_{q_1 q_2}\psi^{1^+,K=0}_{q_1 q_2}{\cal U}^+_{q_1 q_2}|0\rangle .
\end{eqnarray}
The amplitudes $\psi^{1^+,K=1}_{q_1 q_2}$ and $\psi^{1^+,K=0}_{q_1 q_2}$ are normalized.
For compactness, we do not show in (\ref{Eq26}) and (\ref{Eq27}) of the operators ${\bar A}^+_{qq'}$
and ${\bar {\cal U}}^+_{qq'}$, although their presence is implied. The terms with the backward  amplitudes 
$\varphi^{1^+ K=1}_{qq'}$ and $\varphi^{1^+ K=0}_{qq'}$ multiplied by the two-quasiparticle annihilation
operators are not included in (\ref{Eq26}) and (\ref{Eq27}) because of the smallness of their effects.
The final expression for the Coriolis mixing matrix element is
\begin{eqnarray}
\label{Eq28}
\langle 1^+ K=1| H_{Coriolis} |1^+ K=0\rangle =\nonumber\\
4\frac{\hbar^2}{\Im}\sum_{q_1 q_2 q}\psi^{1^+ K=1}_{q_1 q_2}\langle q_2, \sigma=+1 |j_x|q,\sigma =+1\rangle v^{(+)}_{q_2 q}\psi^{1^+ K=0}_{q q_1}
\end{eqnarray}
where $\langle q_2, \sigma=+1 |j_x|q,\sigma =+1\rangle$ is the matrix element of the single particle intrinsic angular momentum operator, $v^{(+)}_{q_2 q}=u_{q_2}u_q + v_{q_2}v_q$, and $u-v$ are Bogoliubov transformation coefficients.

Below we consider also Coriolis mixing of the two-phonon states. However, calculation of the corresponding matrix elements is reduced to calculation of the matrix elements between the one-phonon states.

\section{Results}

Calculations are carried out for $^{156,158,160}$Gd and $^{160,162,164}$Dy. The results of calculations are presented in Table 1. As it is seen from Table 1 the calculated values of the Coriolis matrix elements fluctuate significantly from one pair of states to the other pair. It is likely that the relatively stable
value of the Coriolis matrix element would be expected in the case of mixing of the collective states. However,
scissor $K^{\pi}=1^+$ state is quite fragmented. The structure of the $K^{\pi}=0^+$ states is also changes with excitation energy. The values of the calculated Coriolis matrix elements vary from 1 keV to 200 keV.
In many of the considered cases the Coriolis matrix elements are comparable with an energy distances between
the mixed levels. In average, the calculated matrix elements are of the same order as those obtained in
\cite{Scholten}.

\begin{table}
\caption{Coriolis mixing matrix elements between the magnetic type $K^{\pi}=1^+$ and the nearby
lying $K^{\pi}=0^+$ excited states.}
\begin{tabular}{cccc}
\hline
Nucleus & E($K^{\pi}=1^+$) (keV) & E($K^{\pi}=0^+$) (keV) & $|\langle 1^+,K=1|H_{Coriolis}|1^+,K=0\rangle |$ (keV)\\
\hline
$^{156}$Gd  & 2790 & 2750 & 53 \\
$^{156}$Gd  & 2790 & 2800 & 68 \\
$^{156}$Gd	& 2790 & 2880 & 26 \\
$^{156}$Gd  & 3110 & 2750 & 209 \\
$^{156}$Gd  & 3110 & 2800 & 29  \\
$^{156}$Gd  & 3110 & 2880 & 82  \\
\hline
$^{158}$Gd  & 3220 & 3000 & 39  \\
$^{158}$Gd  & 3220 & 3130 & 11  \\
$^{158}$Gd  & 3220 & 3190 & 112  \\
$^{158}$Gd  & 3220 & 3230 & 5   \\
$^{158}$Gd  & 3270 & 3000 & 29  \\
$^{158}$Gd  & 3270 & 3130 & 18  \\
$^{158}$Gd  & 3270 & 3190 & 24  \\
$^{158}$Gd  & 3270 & 3230 & 25  \\
\hline
$^{160}$Gd  & 2710 & 2760 & 52  \\
$^{160}$Gd  & 2710 & 3060 & 15  \\
$^{160}$Gd  & 3020 & 2760 & 14  \\
$^{160}$Gd  & 3020 & 3060 & 78  \\
\hline
$^{160}$Dy  & 2820 & 2840 & 1   \\
$^{160}$Dy  & 2820 & 3020 & 1   \\
$^{160}$Dy  & 2930 & 2840 & 1   \\
$^{160}$Dy  & 2930 & 3020 & 60  \\
\hline
$^{162}$Dy  & 3090 & 2870 & 2   \\
$^{162}$Dy  & 3090 & 2930 & 35  \\
$^{162}$Dy  & 3200 & 2870 & 5   \\
$^{162}$Dy  & 3200 & 2930 & 1   \\
\hline
$^{164}$Dy  & 3070 & 3200 & 9   \\
$^{162}$Dy  & 3230 & 3200 & 3   \\
\hline
\end{tabular}
\end{table}

It is also seen from Table 1 that the calculated matrix elements for Gd isotopes are larger in average than for Dy isotopes. It indicates that Gd isotopes could be more suitable for finding deviations from Alaga rule
in M1 transitions from the 1$^+$ states to the states of the ground band.

\section{Conclusion}
We have calculated the Coriolis matrix elements for mixing of the $1^+ K=1$
and $1^+ K=0$ states in several Gd and Dy isotopes. The results obtained show that the values
of the matrix elements fluctuate significantly. In many cases the Coriolis matrix elements are equal
to several tens of keV. The results obtained indicate that Gd isotopes could be more suitable for
finding deviations from Alaga rule in M1 transitions from the 1$^+$ states to the states of the ground band.

\section{Acknowledgments}
The authors would like to thank Prof. N.Pietralla for helpful discussions. The authors acknowledge support by the Ministry of Education and Science (Russia) under Grant No. 075-10-2020-117.

\section{Data Availability Statement}
This manuscript has no associated data or the data will not be deposited. [Authors' comment: All the relevant and concerned data has been provided in the manuscript.]

\end{document}